\documentclass[10pt, a4paper, twocolumn]{IEEEtran}

\usepackage{amsmath,epsfig,amssymb,verbatim,amsopn,cite,subfigure,multirow}
\usepackage{balance,nopageno}
\usepackage{algorithm,algorithmic}
\usepackage[usenames,dvipsnames]{color}
\usepackage[all]{xy}  
\usepackage{url}
\usepackage{amsfonts}
\usepackage{amssymb}
\usepackage{epsfig}
\usepackage{epstopdf}
\usepackage{slashbox}
\usepackage{bm}
\usepackage{balance}
\usepackage[margin=18.0mm,top=18mm,bottom=18mm]{geometry}

\newtheorem{Lemma}{Lemma}

\newtheorem{Corollary}[Lemma]{Corollary}

\newtheorem{proposition}{Proposition}

\def\ie{{i.e}.}

\newcommand{\E}[1]{\mathbb{E}\left\{#1\right\}}

\newcommand{\qf}{{\bf f}}

\newcommand{\qh}{{\bf h}}

\newcommand{\qn}{{\bf n}}

\newcommand{\qw}{{\bf w}}

\newcommand{\qB}{{\bf B}}

\newcommand{\qH}{{\bf H}}
\newcommand{\qI}{{\bf I}}

\def\ie{\emph{i.e}.}

\newcommand{\Snr}{\sigma_{R}^2}
\newcommand{\Sn}{N_0}
\newcommand{\Snu}{\sigma_{n_{1}}^2}
\newcommand{\Snuu}{\sigma_{n_{2}}^2}
\newcommand{\Ps}{P_S}
\newcommand{\Pre}{P_R}

\newcommand{\Nrx}{N_{\mathsf{R}}}
\newcommand{\Ntx}{N_{\mathsf{T}}}
\newcommand{\SUu}{U_{1,i}}
\newcommand{\SUun}{U_{1,i}^*}
\newcommand{\SUr}{\mathbb{R}}
\newcommand{\SUuu}{U_{2,i}}
\newcommand{\SUuun}{U_{2,i}^*}

\newcommand{\MRC}{\mathsf{MRC}}

\newcommand{\ZF}{\mathsf{ZF}}

\newcommand{\PoutuTZF}{\mathsf{P_{out,1}^{TZF}}}
\newcommand{\PoutuTZFn}{\mathsf{P_{out,1^*}^{TZF}}}

\newcommand{\PoutuTZFU}{\mathsf{P_{out,1}^{TZF,U}}}
\newcommand{\PoutuTZFUn}{\mathsf{P_{out,1^*}^{TZF,U}}}

\newcommand{\PoutuuTZF}{\mathsf{P_{out,2}^{TZF}}}
\newcommand{\PoutuuTZFn}{\mathsf{P_{out,2^*}^{TZF}}}

\newcommand{\Poutu}{\mathsf{P_{out,1}}}
\newcommand{\Poutuu}{\mathsf{P_{out,2}}}

\newcommand{\Prob}{\textnormal{Pr}}

\newcounter{mytempeqcounter}

\newcommand{\bigformulatop}[2]{%
  \begin{figure*}[!t]
    \normalsize
    \setcounter{mytempeqcounter}{\value{equation}}
    \setcounter{equation}{#1}
    #2

    \setcounter{equation}{\value{mytempeqcounter}}
    \hrulefill
    \vspace*{4pt}
  \end{figure*}
}

\newcommand{\ThankFour}{The work of Z. Ding was supported by the UK EPSRC under grant number EP/N005597/1. The work of Z. Ding and H. A. Suraweera were supported by H2020-MSCA-RISE-2015 under grant number 690750.}

\title{Full-duplex Multi-Antenna Relay Assisted Cooperative Non-Orthogonal Multiple Access}

\author{{Zahra Mobini$^\dag$, Mohammadali Mohammadi$^\dag$, Himal A. Suraweera$^\ddag$, Zhiguo Ding$^\S$ }\\
\small{$^\dag$Faculty of  Engineering, Shahrekord University, Iran\\
$^\ddag$Department of Electrical and Electronic Engineering, University of Peradeniya, Sri Lanka\\
$^\S$School of Computing and Communications, Lancaster University, United Kingdom \\
Email: \{z.mobini, m.a.mohammadi\}@eng.sku.ac.ir, himal@ee.pdn.ac.lk, z.ding@lancaster.ac.uk}}\normalsize

\begin{document}
\maketitle
\thispagestyle{empty}

\begin{abstract}
We consider a cooperative non-orthogonal multiple access (NOMA) network in which a full-duplex (FD) multi-antenna relay assists transmission from a base station (BS) to a set of far users with poor channel conditions, while at the same time the BS transmits to a set of near users with strong channel conditions. We assume imperfect self-interference (SI) cancellation at the FD relay and imperfect inter-user interference cancellation at the near users. In order to cancel the SI at the relay a zero-forcing based beamforming scheme is used and the corresponding outage probability analysis of two user selection strategies, namely random near user and
random far user (RNRF), and nearest near user
and nearest far user (NNNF), are derived. Our finding suggests that significant performance improvement can be achieved by using the FD multi-antenna relay compared to the counterpart system with a half-duplex relay. The achieved performance gain depends on network parameters such as the user density, user zones, path loss and the strength of the inter-user interference in case of near users. We also show that the NNNF strategy exhibits a superior outage  performance compared to the RNRF strategy, especially in the case of near user.\let\thefootnote\relax\footnotetext{\ThankFour}
\end{abstract}

\section{Introduction}\label{sec:intro}
Due to rapid increase in services with ever growing bandwidth demand and higher quality of service requirement of wireless subscribers and networks, spectral efficiency is considered as a main challenge affecting the design of emerging wireless networks. To enhance spectral efficiency,  non-orthogonal multiple access (NOMA) concept aims to realize multiple access in the power domain and accordingly serve multiple users in the same time and frequency~\cite{Saito:VTC2013,zhiguo:2017:MCOM}.  NOMA achieves high spectral efficiency by squeezing a user with a strong channel condition, hereafter called the near user, into the spectrum occupied by a user with a poor channel condition, hereafter called the far user. However, the benefits of NOMA cannot be reaped without having some cost increment. Particularly, the reliability of the far users may adversely affected due to the fact that the near users co-exist with the far users~\cite{zhiguo:2017:MCOM}.
Recently some efforts have been made to exploit cooperative techniques in NOMA systems as an efficient way to improve the performance of the far users. The existing cooperative NOMA systems can be classified into user-assisted cooperative NOMA and relay-assisted cooperative NOMA system. For user-assisted cooperative NOMA, the near user, helps the far user, exploiting the fact that the near user is able to decode the information for both users~\cite{Zhiguo:CLET:2015,Zhiguo:JSAC:2016}. For the relay-assisted cooperative NOMA, a dedicated relay is employed to assist the far user~\cite{Men:JCOML:2015,KIM:2015:JCOML,Ding:2016:wcoml,Liang:CLET}. The authors in~\cite{Zhiguo:JSAC:2016} investigated the application of wireless power transfer in a user-assisted cooperative NOMA network where the user positions  were modeled using stochastic geometry.
In~\cite{Men:JCOML:2015} a dedicated relay has been used in to serve multiple users equipped with multiple antennas. The authors in~\cite{KIM:2015:JCOML}, have shown that remarkable performance gains in terms of the outage performance and sum capacity can be obtained through cooperation in a relay-assisted NOMA system. The impact of relay selection on the performance of relay-assisted cooperative NOMA system  was examined in~\cite{Ding:2016:wcoml}. In~\cite{Liang:CLET} the outage performance for a downlink cooperative NOMA scenario with the help of a relay is investigated.

Common to all the above works~\cite{Men:JCOML:2015,KIM:2015:JCOML,Ding:2016:wcoml,Liang:CLET} is the assumption of half-duplex (HD) operation at the relaying node. However, the implementation of HD relaying requires additional time resources, which results in a loss of the spectral efficiency. Thus, in an effort to recover the spectral loss, full-duplex (FD) relaying,  where the relay receives and transmits simultaneously in the same frequency band could be utilized~\cite{Ashutosh:JSAC:2014,Sabharwal:TWC2012}. However, FD operation in a relay-assisted NOMA system introduces several challenges such as inter-user interference at near users due to relay transmission to far user and self-interference (SI) at the FD relay due to signal leakage from the relay's output to the input.  Nevertheless, many effective SI cancellation methods have been proposed~\cite{Riihonen:JSP:2011, Himal:2014:JSAC} to enhance the practical application of FD implementation. As such, a few recent studies propose the combination of FD operation and the NOMA principle~\cite{Ding:TVT:2017,Caijun:CLET:2016,Ding:2017:ICC}. In~\cite{Ding:TVT:2017}, an FD device-to-device aided cooperative NOMA scheme was proposed where the near user is FD capable and assist the base station (BS) transmissions to the far user. In~\cite{Caijun:CLET:2016}, an FD relay-assisted cooperative NOMA with dual-users was examined. It was shown that, the proposed FD relay-assisted NOMA system in~\cite{Caijun:CLET:2016} achieves better performance compared to the HD one in the low to moderate signal-to-noise ratio (SNR) regimes. The authors in~\cite{Ding:2017:ICC} provided the diversity analysis of a hybrid FD/HD user-assisted NOMA system with two users.


The above works on FD NOMA systems have considered a single antenna relay. However, recent studies have shown that SI cancellation can be performed effectively in spatial domain, if multiple antennas are used~\cite{Riihonen:JSP:2011,Mohammadi:TCOM:2016}. However, recently stochastic geometry has been widely used to model user location with high accuracy. Therefore, it is important to study the FD multi-antenna relay-assisted cooperative NOMA in a network under random spatial deployment of users. We assume a NOMA system with a BS, relay and two sets of users and use stochastic geometry to capture the impact of random locations of the users in a cooperative NOMA system. In particular, we consider a network with two groups of randomly deployed users: near users, deployed within a disc, and far users, deployed within a ring, where their locations are modeled as homogeneous Poisson point processes (PPPs). We consider the following user selection strategies, namely (i) random near user and random far user (RNRF) selection; (ii) nearest near user and nearest far user (NNNF) selection. We exploit an FD multi-antenna relay and employ maximal ratio combining (MRC) at the relay input and zero forcing (ZF) at the relay output, to obtain receive and transmit beamformers with the objective of cancelling the SI at the relay. Our main contributions are summarized as follows:
\begin{itemize}
\item We consider a realistic scenario with imperfect SI cancellation at the relay and imperfect inter-user interference cancellation at the near users and derive outage probability expressions for the RNRF and NNNF strategies. In addition, to extract insights and highlight the system behavior, closed-form upper bound and lower bound on the outage probability of the near users are presented.
\item Our findings reveal that compared to the RNRF strategy, NNNF provides a superior  outage probability for both the near and far users. Moreover, comparing the proposed FD cooperative NOMA system with the HD counterpart, we confirm that the FD cooperative NOMA achieves a better outage performance.
\end{itemize}
\emph{Notation:}  We use bold upper case letters to denote matrices, bold lower case letters to denote vectors. The superscripts  $(\cdot)^{T}$, $(\cdot)^{*}$, $(\cdot)^{\dag}$ stand for transpose, conjugated,  and conjugate transpose respectively; the Euclidean norm of the vector and the expectation are denoted by $\|\cdot\|$ and $\mathbb{E}\left\{\cdot\right\}$ respectively; $\Prob(\cdot)$ denotes the probability; $f_X(\cdot)$ and $F_X(\cdot)$ denote the probability density function (pdf) and cumulative distribution function (cdf) of the random variable (RV) $X$, respectively; $\mathcal{CN}(\mu,\sigma^2)$ denotes a circularly symmetric complex Gaussian RV $x$ with mean $\mu$ and variance $\sigma^2$;  $\Gamma(a,x)$ is upper incomplete Gamma function; $\gamma(a,x)$ is lower incomplete Gamma
function~\cite[Eq. (8.350)]{Integral:Series:Ryzhik:1992}.
\section{System Model}
\begin{figure}[t]
\centering
\vspace{0.9em}
\includegraphics[width=60mm, height=60mm]{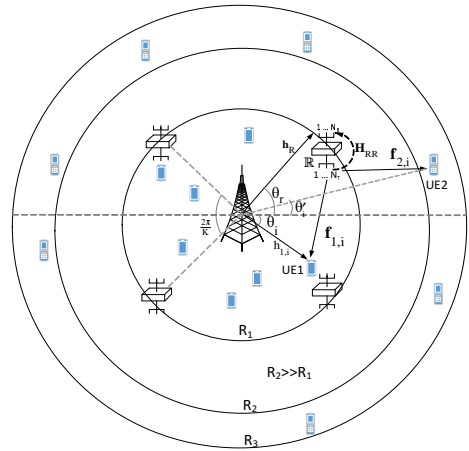}
\vspace{-0.8em}
\caption{An illustration of relay-assisted cooperative NOMA.
The spatial distributions of the near users and the far users
follow homogeneous PPPs. FD multi-antenna relays are symmetrically deployed
at a distance $R_1$ from the BS in a circular fashion. }
\vspace{-1em}
\label{fig: system model}
\end{figure}

Consider a network with a BS and two groups of randomly deployed users: near and far users as shown in Fig.~\ref{fig: system model}~\cite{Zhiguo:JSAC:2016}. We assume that the near users $\{U_{1,i}\}$, $i=1,\cdots,N_{U_1}$, are deployed within disc with radius $R_1$, denoted by $D_n$, and the far users $\{U_{2,i}\}$, $i=1,\cdots,N_{U_2}$, are deployed within ring with radius $R_2$ and $R_3$, denoted by $D_f$, (assuming $R_2\gg R_1$). The locations of the near and far users are modeled as homogeneous Poisson point processes (PPPs) $\Phi_n$ and $\Phi_f$, respectively, with the densities $\lambda_n$ and $\lambda_f$. Without loss of generality, we focus on the NOMA with two users: one from near and one from far user sets where the near user $U_{1,i}$ directly communicates with the BS. However, as in~\cite{Caijun:CLET:2016,KIM:2015:JCOML} we assume that there is no direct link between the BS and the far user $\SUuu$. Therefore, we exploit $K$ fixed decode-and-forward (DF) relays, $\{\mathbb{R}_k\}$, $k=1,\cdots,K$, symmetrically deployed at a distance $R_1$ from the cell center in a circular fashion, that forward the information to the far users. Similar to~\cite{Zhiguo:JSAC:2016,Caijun:CLET:2016,Liang:CLET} we assume a single antenna BS communication assisted by an FD multi-antenna relay, where the total number of antennas at the FD relay is $N = \Ntx + \Nrx$ of which $\Nrx$ antennas are dedicated for reception and $\Ntx$ antennas are used for transmission. For a more realistic propagation model, we assume that the links experience both large-scale path loss effects and small-scale  fading. All channels are assumed to be quasi-static Rayleigh fading, which means that the channel coefficients are constant
over the block time  $T$ and vary independently between different blocks. Thus, each element of these complex fading channel coefficients are circular symmetric complex Gaussian random variables. We assume that $\alpha\geq 2$ is the path loss exponent and $\ell( x, y ) =\|x - y \|$ is the Euclidean distance between two nodes. If $y$ is the origin, the index $y$ will be omitted, i.e. $\ell(x, 0) = \ell(x)$.
Before the transmission, the two users $\SUu$ and $\SUuu$ are selected to perform NOMA with the aid of the selected relay, denoted by $\mathbb{R}$, where the selection criterion for user selection and relay selection will be discussed in Subsection~\ref{sec:relay selection}.
\vspace{-0.0em}
\subsection{Transmission Protocol}
According to the NOMA concept, BS transmits a combination of messages to both users and the selected relay $\SUr$ as
\vspace{-0.0em}
\begin{align}
s[n]=\sqrt{\Ps a_{1,i}}x_{1,i}[n]+\sqrt{\Ps a_{2,i}}x_{2,i}[n],
\end{align}
where $\Ps$ is the transmit power of the BS and $x_{k,i}, k\in\{1,2\}$ denotes the information symbol to $U_{k,i}$, and $a_{k,i}$ denotes the power allocation coefficient, such that $a_{1,i} + a_{2,i} =1$ and $a_{1,i}< a_{2,i}$.
The selected relay $\SUr$ operates in the FD mode and hence it simultaneously receives $s[n]$ and forwards $r[n]$ to the $\SUuu$. The received signal at $\SUr$ is given by
\vspace{-0.5em}
\begin{align}
y_R[n]=R_1^{-\frac{\alpha}{2}} \qh_R s[n]+ \qH_{RR} r[n]+\qn_R[n],
\end{align}
where $\qh_R\in\mathcal{C}^{\Nrx\times 1}$ is the channel between the source and relay and its entries follow identically independent distributed (i.i.d), $\mathcal{CN}(0,1)$,
$r[n]$ is the transmitted relay signal satisfying ${\mathbb{E}}\left\{r[n]r^{\dag}[n]\right\}=P_R$, given by
\begin{align}
r[n] =\sqrt{\Pre}\qw_{t,i} x_{2,i}[n-\delta],
\end{align}
$\delta$ accounts for the time delay caused by relay processing, and $\qn_R[n]$ is the AWGN at the relay with $\E{\qn_R\qn_R^\dag}=\Snr\qI$.  We model the $N_R \times N_T$ residual SI channel $\qH_{RR}$  as i.i.d $\mathcal{CN}(0,\sigma^2_{RR})$ RVs~\cite{Sabharwal:TWC2012,Riihonen:JSP:2011}. Since the relay $\SUr$ adopts the DF protocol, upon receiving the signal, it first  applies a linear combining vector $\qw_r$ on $y_R$ to obtain an estimate of $s$, denoted by $\hat s[n]$, as
\begin{align}
 \hat s[n] =  R_1^{-\frac{\alpha}{2}} \qw_r^{\dag}\qh_{R}s[n] + \qw_r^{\dag}\qH_{RR} r[n]\! +\!\qw_r^{\dag}\qn_R[n].
\end{align}
Then it decodes the information intended for $\SUuu$ while treating the symbol of $\SUu$ as interference. Finally, it forwards $x_{2,i}[n-\delta]$ to $\SUuu$ using the transmit beamforming vector $\qw_{t,i}$. Let $\|\qw_{t,i}\|^2=\|\qw_r\|^2=1$ and the received signal-to-interference-plus-noise ratio  (SINR) at the selected relay $\SUr$  is given by
\vspace{-0.0em}
\begin{align}\label{eq:gammaR}
\gamma_R=\frac{\rho_sa_{2,i}R_1^{-\alpha}|\qw_r^\dag \qh_R|^2}{\rho_sa_{1,i}R_1^{-\alpha}|\qw_r^\dag \qh_R|^2+\rho_r|\qw_r^\dag \qH_{RR} \qw_{t,i}|^2+1},
\end{align}
where $\rho_s=\frac{P_S}{\Sn}$ and $\rho_r=\frac{P_R}{\Sn}$ (without lost of generality, it is assumed that the mean power of noise at all users and relay is the same and denoted by$\Sn$).
On the other hand, the received signal at $\SUu$ can be written as
\vspace{-0.0em}
\begin{align}\label{eq:y_{1,i}}
&y_{1,i}[n]=\ell(\SUu)^{\frac{-\alpha}{2}}h_{1,i}s[n]+\nonumber\\
&\qquad\ell(\SUr,\SUu)^{\frac{-\alpha}{2}}\sqrt{\Pre} \qf_{1,i}^T \qw_{t,i} x_{2,i}[n-\delta]+n_{1,i}[n],
\end{align}
where $h_{1,i} \sim \mathcal{CN}(0,1)$ is the channel between the source and $\SUu$, $\qf_{1,i}\in\mathcal{C}^{N_T\times 1}$ denotes the channel between the relay and $\SUu$,  and $n_{1,i}[n]\sim\mathcal{CN}(0,\Snu)$ denotes the additive white Gaussian noise (AWGN) at the $\SUu$. Moreover, $\ell(\SUr,\SUu)=\sqrt{R_1^2+\ell(\SUu)^2-2R_1\ell(\SUu)\cos(\theta_r-\theta_{i})}$, where $\theta_r$ denotes the
angle of the selected relay $\SUr$ from reference x-axis, and $\theta_{i}$ denotes the angle of the $\SUu$ from reference x-axis. Moreover, $-\pi\leq\theta_r-\theta_{i}\leq\pi$. Applying the principle of NOMA concept, successive interference cancellation (SIC) is carried out at $\SUu$. In particular, $\SUu$ first decodes the message of $\SUuu$, \ie, $x_{2,i}$, then subtracts it from the received signal to detect its own message~\cite{Saito:VTC2013}. Therefore, the received SINR at $\SUu$ to detect $x_{2,i}$ of $\SUuu$ is given by
\vspace{-0.0em}
\begin{align}\label{eq:gamma12i}
\gamma_{1,i}^{x_{2,i}}&=\\
&\frac{\rho_sa_{2,i}\ell(\SUu)^{-\alpha}|h_{1,i}|^2}{\rho_sa_{1,i}\ell(\SUu)^{-\alpha}|h_{1,i}|^2\!+\!\rho_r\ell(\SUr,\SUu)^{-\alpha}|\qf_{1,i}^T \qw_{t,i}|^2\!+\!1}, \nonumber
\end{align}
and the received SINR at $\SUu$ to detect $x_{1,i}$  is given by
\begin{align}\label{eq:gamma11i}
\gamma_{1,i}^{x_{1,i}}=\frac{\rho_sa_{1,i}\ell(\SUu)^{-\alpha}|h_{1,i}|^2}{\rho_r\ell(\SUr,\SUu)^{-\alpha}|\qf_{1,i}^T \qw_{t,i}|^2+1}.
\end{align}
According to the principle of NOMA, $x_{2,i}[n-\delta]$ is priory known to $\SUu$ and thus $\SUu$ can remove it via interference cancellation~\cite{Caijun:CLET:2016}. Nevertheless, here, we consider a realistic imperfect interference cancellation wherein $\SUu$ cannot perfectly remove $x_{2,i}[n-\delta]$. In particular, we model $\qf_{1,i} \sim \mathcal{CN}(0,q_r)$ as the residual inter-user interference channel where the parameter $q_r$ presents the strength of inter-user interference~\cite{Caijun:CLET:2016}. Specifically, $q_r=0$ implies perfect interference cancellation at $\SUu$.

Finally, the observation at $\SUuu$ can be expressed as follows:
\begin{align}
y_{2,i}[n]\!=\!\sqrt{\Pre} \ell(\SUr,\SUuu)^{\frac{-\alpha}{2}}\qf_{2,i}^T\qw_{t,i}x_{2,i}[n\!-\!\delta]\!+\!n_{2,i}[n],
\end{align}
where $\ell(\SUr,\!\SUuu)\!=\!\sqrt{\!R_1^2\!+\!\ell(\SUuu)^2\!-\!2R_1\ell(\SUuu)\!\cos(\theta_r\!-\!\acute{\theta}_{i})}$, $\acute{\theta}_{i}$ denotes  the angle of $\SUuu$ from reference x-axis, $\qf_{2,i}\in\mathcal{C}^{\Ntx\times 1}$ denotes the channel between the $\SUr$ and $\SUuu$ and $n_{2,i}[n]\sim\mathcal{CN}(0,\Snuu)$ denotes the AWGN at the $\SUuu$. Therefore, the received SNR at $\SUuu$  is given by
\begin{align}\label{eq:gamma22i}
\gamma_{2,i}^{x_{2,i}}=\rho_r \ell(\SUr,\SUuu)^{-\alpha}|\qf_{2,i}^T\qw_{t,i}|^2.
\end{align}
\subsection{Beamforming Scheme}
In this subsection, we design receive and transmit beamformers at the selected relay $\SUr$.
We propose a beamforming solution, namely transmit zero-forcing (TZF), where relay takes advantage of the multiple transmit antennas ($\Ntx>1$) to completely cancel the SI~\cite{Suraweera:TWC:2014}. Moreover, MRC is applied at the relay input, i.e., $\qw_r^{\MRC} = \frac{\qh_R}{\|\qh_R\|}$.
Therefore, the optimal transmit beamforming vector $\qw_{t,i}$ is obtained by solving the following problem:
\begin{align}\label{eqn:wt}
    \max_{\|\qw_{t,i}\|=1} &\hspace{0.9em}  | \qf_{2,i}^T \qw_{t,i}|^2 \nonumber\\
     \mbox{s.t.} &\hspace{1em} \qh_R^\dag\qH_{RR}\qw_{t,i} =0.
\end{align}

The coexistence of near users and far users in the NOMA systems results in performance degradation for the far users with poor channel conditions and hence here we improve the outage probability of them\footnote{As an alternative ZF-based beamforming  solution, $\qw_{t,i}$ can be set using the maximal ratio transmission principle (MRT) and $\qw_{r,i}$ can be designed with the ZF criterion.  The performance analysis of the ZF/MRT scheme is left as a future work.}. Specifically, in~\eqref{eqn:wt} we maximize $| \qf_{2,i}^T \qw_{t,i}|^2$ and consequently the second-hop SNR at the far users.

Given the optimization problem in~\eqref{eqn:wt}, using similar steps as in \cite{Mohammadi:TCOM:2016}, the optimal transmit vector $\qw_{t,i}$
is obtained as $\qw_{t,i}^{\ZF} = \frac{\qB \qf_{2,i}^*}{\|\qB \qf_{2,i}^*\|},$ where $\qB = \qI_{\Ntx} - \frac{  \qH_{RR}^\dag \qh_R \qh_R^\dag \qH_{RR}}{\| \qh_R^\dag \qH_{RR}\|^2}$. Accordingly, substituting the $\qw_r^{\MRC}$ and $\qw_{t,i}^{\ZF}$ into~\eqref{eq:gammaR},~\eqref{eq:gamma12i},~\eqref{eq:gamma11i} and~\eqref{eq:gamma22i}
$\gamma_R$, $\gamma_{1,i}^{x_{2,i}}$, $\gamma_{1,i}^{x_{1,i}}$ and $\gamma_{2,i}^{x_{2,i}}$ can be obtained, respectively.
\vspace{-0.5em}
\subsection{User Selection and Relay Selection Strategies}\label{sec:relay selection}
We consider two user selection strategies namely RNRF and NNNF.
For the RNRF strategy, the BS randomly selects a near user $\SUu$ and a far user $U_{2,i}$ from the two groups of users. For the NNNF strategy, a user within the disc $D_n$ with the shortest distance to the BS is selected as a near user $\SUun$ and the user within ring $D_f$ with the shortest distance to the BS is selected as a far user $\SUuun$.

On the other hand, for each user selection strategy,  the relay with the minimum Euclidean distance from the selected far user is chosen for cooperative NOMA. We can define the relay selection criterion as
\begin{align}\label{eq:reay:selec}
 \min\{\ell(\mathbb{R}_k, U_{2,i}), k\in\{1,\cdots,K\}\}.
 \end{align}
This relay selection strategy is suitable for practical scenarios, wherein the far users are much farther away from the BS in comparison with near users and thus have the poor channel conditions. Accordingly, the criterion in~\eqref{eq:reay:selec} can improve the reception reliability of the far users.
\section{Cooperative NOMA with RNRF Selection }
Now, we aim to characterize the performance of cooperative NOMA with the RNRF user selection. This strategy achieves fairness and it can be performed without the knowledge of the instantaneous channel state information (CSI) of the users.
\bigformulatop{16}{
\begin{align}\label{eq:out:u1:def3}
\PoutuTZF\stackrel{(a)}=&\int_{D_n}\int_{-\pi}^{\pi}\int_{0}^{\infty}\frac{1}{q_r}\left(1-e^{-{\mu\left(\rho_r\ell(\SUr,\SUu)^{-\alpha}y+1\right) \ell(\SUu)^{\alpha}}}\right)e^{-\frac{y}{q_r}}f_{\Theta_i}(\theta_i)f_{W_{n,i}}(w_{n,i})dy d\theta_i dw_{n,i}\nonumber\\
=&\int_{D_n}\int_{-\pi}^{\pi}\left(1-\frac{e^{-\mu\ell(\SUu)^{\alpha}}}{1+q_r\rho_r\mu\ell(\SUr,\SUu)^{-\alpha}\ell(\SUu)^{\alpha}}\right)f_{\Theta_i}(\theta_i)f_{W_{n,i}}(w_{n,i}) d\theta_i dw_{n,i},
\end{align}
}
\subsection{Outage Probability at the Near Users}
An outage event at the near user $\SUu$ happens when it cannot decode $x_{2,i}$ or when it can decode $x_{2,i}$ but cannot decode $x_{1,i}$. Let $\tau_1 = 2^{\mathcal{R}_1}-1$ and $\tau_2 = 2^{\mathcal{R}_2}-1$, where $\mathcal{R}_1$ and $\mathcal{R}_2$ are the transmission rates at $\SUu$ and $\SUuu$, respectively. The outage probability at $\SUu$ can be expressed as
\begin{align}\label{eq:outage event near}
\Poutu &=1-\Prob\left(\gamma_{1,i}^{x_{2,i}}>\tau_2, \gamma_{1,i}^{x_{1,i}}>\tau_1\right),
\end{align}
\begin{proposition}
\label{prob:outage:u1:TZF}
The outage probability of $\SUu$ with the TZF scheme is given by
\begin{align}\label{prob:outage:u1:TZF}
&\PoutuTZF=\frac{1}{\pi R_1^2}\int_{0}^{R_1}\int_{-\pi}^{\pi}\left(1-\right.\\&\left.\frac{e^{-\mu r^\alpha}}{1\!+q_r\rho_r\mu\left(R_1^2+r^2\!-\!2rR_1\cos(\theta_r\!-\theta_i)\right)^{-\frac{\alpha}{2}}r^{\alpha}}
\right)r d\theta_i dr,\nonumber
\end{align}
\end{proposition}
if $\tau_2\leq\frac{a_{2,i}}{a_{1,i}}$, otherwise $\PoutuTZF=1$, where  $\mu={\max\left(\frac{1}{\zeta},\frac{\tau_1}{b_1}\right)}$ with $\zeta=\frac{\rho_sa_{2,i}-\rho_sa_{1,i}\tau_2}{\tau_2}$.
\emph{proof:}
Let $Y_0\triangleq|\qf_{1,i}^T \qw_{t,i}^{\ZF}|^2$, $Y_1=|h_{1,i}|^2$, $b_0=\rho_sa_{2,i}$ and $b_1=\rho_sa_{1,i}$. Applying $\gamma_{1,i}^{x_{2,i}}$, $\gamma_{1,i}^{x_{1,i}}$ in~\eqref{eq:outage event near}, the outage of $\SUu$ can be written as
\begin{align}\label{eq:out:u1:def1}
&\PoutuTZF\!=\!1\!-\!\Prob\!\left(\!\frac{b_0\ell(\SUu)^{-\!\alpha}Y_1}{b_1\ell(\SUu)^{-\!\alpha}Y_1\!+\!\rho_r\ell(\SUr,\SUu)^{\!-\!\alpha}Y_0\!+\!1}\!>\!\tau_2,\right.\nonumber
\\&\qquad\qquad\left.\frac{b_1\ell(\SUu)^{-\alpha}Y_1}{\rho_r\ell(\SUr,\SUu)^{-\alpha}Y_0\!+\!1}>\tau_1\right)\nonumber\\
&\quad=\Prob\left(\rho_r\ell(\SUr,\SUu)^{-\!\alpha}Y_0+1>\frac{1}{\mu}\ell(\SUu)^{-\alpha}Y_1\right).
\end{align}
In~\eqref{eq:out:u1:def1}, if $\tau_2>\frac{a_{2,i}}{a_{1,i}}$, $\mu<0$ and hence $\PoutuTZF=1$. On the other hand, when $\tau_2\leq\frac{a_{2,i}}{a_{1,i}}$ we have
\begin{align}\label{eq:out:u1:tzf:def2}
\PoutuTZF=\Prob\left(Y_1\leq {\left(\rho_r\ell(\SUr,\SUu)^{-\alpha}Y_0+1\right)\mu \ell(\SUu)^{\alpha}}\big| Y_0\right).
\end{align}
For the case $\tau_2\leq\frac{a_{2,i}}{a_{1,i}}$, note that we model the location of
the near and far users as i.i.d. points in $D_n$ and $D_f$, denoted by $W_{n,i}$ and $W_{f,i}$, respectively,  with the pdf $f_{W_{n,i}}(w_{n,i})=\frac{\lambda_n}{\mu_{n}}=\frac{1}{\pi R_1^2}$ and $f_{W_{f,i}}(w_{f,i})=\frac{\lambda_f}{\mu_{f}}=\frac{1}{\pi (R_3^2-R_2^2)}$.
Therefore,~\eqref{eq:out:u1:tzf:def2} can be expressed as~\eqref{eq:out:u1:def3} at the top of the  next page where (a) follows by the fact that  $Y_0$ and $Y_1$ are exponential RVs with the cdfs $F_{Y_0}(y)=1-e^{-y/q_r}$ and $F_{Y_1}(y)=1-e^{-y}$, respectively.  Substituting $f_{\Theta_i}(\theta_i)=\frac{1}{2\pi}$ and $f_{W_{n,i}}(w_{n,i})$ into~\eqref{eq:out:u1:def3} we get the desired result in~\eqref{prob:outage:u1:TZF}. $\blacksquare$

The integral in~\eqref{prob:outage:u1:TZF} does not admit a closed-form solution. Nevertheless, it can be solved numerically using popular software packages such as MATLAB. Moreover, we now present approximate closed-form expressions for the lower bound and upper bound of the outage probability. In particular, by setting $\cos(\theta_r-\theta_i)=+1$, $\ell(\SUr,\SUu)$ is minimized and hence the inter-user interference at $\SUu$ is maximized which minimizes $\gamma_{1,i}^{x_{1,i}}$ and $\gamma_{1,i}^{x_{2,i}}$. On the other hand, $\cos(\theta_r-\theta_i)=-1$ results in the minimum inter-user interference at $\SUu$.
\begin{Corollary}
\label{cor:upp:outage:u1:TZF}
The outage probability of $\SUu$ with the TZF scheme can be approximately upper bounded (lower bounded) in closed-form as
\setcounter{equation}{17}
\begin{align}\label{eq:out_tzf_near asymp_upper}
&\PoutuTZFU\approx\frac{\pi}{2M}\sum_{m=1}^{M}\sqrt{(1-\phi_m^2)}\left(1-\right.\\&\left.\frac{e^{-\mu c_m^\alpha}}{1+q_r\rho_r\mu \left(R_1^2+c_m^2-2\eta R_1c_m\right)^{-\frac{\alpha}{2}}c_m^\alpha}\right)(\phi_m+1),\nonumber
\end{align}
where $\eta=1$ (In case of the lower bound, $\eta=-1$),  $c_m=(\phi_m+1)\frac{R_1}{2}$, $\phi_m=\cos(\frac{2m-1}{2M}\pi)$ and $M$ is a parameter to guarantee a desirable complexity-accuracy tradeoff.
\end{Corollary}
\emph{proof:}
By setting $\cos(\theta_r-\theta_i)=1$ in~\eqref{prob:outage:u1:TZF}, the upper bound on the outage probability of $\SUu$ written as
\begin{align}
\label{eq:out_tzf_near lower1}
\PoutuTZFU&=\frac{2}{R_1^2}\\
&
\times \int_{0}^{R_1}\!\!\!\left(\!1\!-\!\frac{e^{-\mu r^{\alpha}}}{1\!+\!q_r\rho_r\mu  (R_1^2\!+\!r^2\!-\!2\eta R_1r)^{\frac{-\alpha}{2}}r^\alpha}\!\right)r dr\!,\nonumber
\end{align}
where $\eta=1$ (In case of the lower bound, $\eta=-1$).
To the best of our knowledge, the integral in~\eqref{eq:out_tzf_near lower1} does not admit a closed-form solution, however by following a similar approach as in~\cite{Zhiguo:JSAC:2016}, we use Gaussian-Chebyshev quadrature~\cite{num:book:hildebrand}
to arrive at~\eqref{eq:out_tzf_near asymp_upper}. $\blacksquare$

\bigformulatop{30}{
\begin{align}\label{eq:outage U1:TZF:TOTAL:nnnf}
\PoutuuTZFn\approx1-\frac{b_4}{\Gamma(\Nrx)}\Gamma\!\left(\!\Nrx,\!\frac{\tau_2R_1^\alpha}{\rho_s(a_{2,i}\!-\!\tau_2a_{1,i})}\!\right)
\sum_{k=0}^{\Ntx-2}\frac{\beta^{k}}{k!}
 \left(\sum_{m=1}^{M}\sqrt{1\!-\!\phi_m^2}s_m^{\alpha k+1}e^{-\left(\beta s_m^\alpha+\pi \lambda_f s_m^2 \right)}\right),
\end{align}
}
\subsection{Outage Probability at the Far Users}
The outage event at $\SUuu$ results due to the following two cases: 1) $\SUr$ cannot decode $x_{2,i}$ and 2) $\SUr$ can decode $x_{2,i}$ but $x_{2,i}$ cannot be decoded correctly by $\SUuu$. Therefore, the outage probability at $\SUuu$ can be written as
\begin{align}\label{eq:outage u2}
\Poutuu\! =\! \Prob\left( \gamma_{R}\!<\!\tau_2\right)\!+\!\Prob\left( \gamma_{R}\!>\!\tau_2\right)\Prob\left( \gamma_{2,i}^{x_2,i}<\tau_2\right)
\end{align}
The following proposition presents the outage probability for an arbitrary choice of $\alpha$.
\begin{proposition}
\label{prob:outage:u2:TZF}
The outage probability of $\SUuu$ with the TZF scheme is given by
\begin{align}\label{eq:outage U1:TZF:TOTAL}
\hspace{-0.25em}\PoutuuTZF&\!=\! 1-\frac{b_3}{\Gamma(\Nrx)}\Gamma\!\!\left(\!\Nrx,\!\frac{\tau_2R_1^\alpha}{\rho_s(a_{2,i}\!-\!\tau_2a_{1,i})}\!\right)
\nonumber\\
&\times{\sum_{k=0}^{\Ntx-2}\frac{\beta^{k-\epsilon}}{k!\alpha}\left(\Gamma\left(\epsilon,\beta R_2^\alpha\right)\!-\!\Gamma\left(\epsilon,\beta R_3^{\alpha}\right)\right)},
\end{align}
where $b_3=\frac{2}{(R_3^2-R_2^2)}$, $\epsilon=k+\frac{2}{\alpha}$ and $\beta=\frac{\tau_2}{\rho_r}$.
\end{proposition}
\emph{proof:}
Let us denote  $Y_2=\rho_sR_1^{-\alpha}\|\qh_R\|^2$ and $Y_3=\|\tilde{\qf}_{2,i}\|^2$. Substituting $\gamma_{R}$ and $\gamma_{2,i}^{x_2,i}$ into~\eqref{eq:outage u2},  $\PoutuuTZF$ can be written as
\begin{align}\label{eq:outage U1:TZF:TOTAL1}
\PoutuuTZF=&\Prob\left(\frac{a_{2,i}Y_2}{a_{1,i}Y_2+1}<\tau_2\right)+
\Prob\left(\frac{a_{2,i}Y_2}{a_{1,i}Y_2+1}>\tau_2\right)\times\nonumber\\
&~\Prob\left(\rho_r\ell(\SUr,\SUuu)^{-\alpha}Y_3 <\tau_2\right).
\end{align}
The RV $Y_2$ follows the chi-square distribution with $2\Nrx$ degrees-of-freedom (DoF). Moreover, to guarantee the implementation of NOMA, the condition $\frac{a_{2,i}}{a_{1,i}}\geq\tau_2$ should be satisfied. Hence, $\PoutuuTZF$ can be written as
\begin{align}\label{eq:outage U1:TZF:TOTAL2}
&\PoutuuTZF=1-\frac{1}{\Gamma(\Nrx)}\Gamma\left(\Nrx,\frac{\tau_2R_1^\alpha}{\rho_s(a_{2,i}-\tau_2a_{1,i})}\right)\!+\!\frac{1}{\Gamma(\Nrx)}\!\times\!\nonumber\\
&\Gamma\left(\Nrx,\frac{\tau_2R_1^\alpha}{\rho_s(a_{2,i}-\tau_2a_{1,i})}\right)
\Prob\left(\rho_r\ell(\SUr,\SUuu)^{-\alpha}Y_3 <\tau_2\right).
\end{align}
The next step is to compute  $\Prob\left(\rho_r\ell(\SUr,\SUuu)^{-\alpha}Y_3 <\tau_2\right)$ where $Y_3$ is an $(\Ntx-1) \times 1$ vector and it follows the chi-square distribution with $2(\Ntx-1)$ DoF.
Moreover, since $R_2\gg R_1$, we can approximate $\ell(\SUr,\SUuu)\approx\ell(\SUuu)$~\cite{Zhiguo:JSAC:2016}. Accordingly, we have
\begin{align}\label{eq:outage U1:TZF:TOTAL3}
&\Prob\left(Y_3 <\frac{\tau_2}{\rho_r\ell(\SUuu)^{-\alpha}}\right)= \nonumber\\
&\qquad1- b_3\sum_{k=0}^{\Ntx-2}\frac{1}{k!}\beta^{k}\int_{R_2}^{R_3} r^{\alpha k+1} e^{-\beta r^\alpha}  dr.
\end{align}
Then, by applying the integral identity~\cite[Eq. (2.33.10)]{Integral:Series:Ryzhik:1992} the desired result in~\eqref{eq:outage U1:TZF:TOTAL} can be obtained. $\blacksquare$

The following corollary provides the outage probability of the far users for the special case $\alpha=2$ and valid in the high SNR regime when the noise term is neglected.
\begin{Corollary}
\label{prob:outage:u2:TZF_alpha2}
In the high SNR regime and for the special case $\alpha=2$,  the outage probability of $\SUuu$ can be simplified as
\begin{align}\label{eq:outage U1:TZF:TOTAL4_alpha2}
\PoutuuTZF=
 {1-\frac{b_3}{2} \sum_{k=0}^{\Ntx-2}\beta^{k}\left(G(R_2)-G(R_3)\right)},
\end{align}
where $G(x)=e^{-\beta x^2}\sum_{j=0}^{k}\frac{x^{2j}}{j!\beta^{k+1-j}}$.
\end{Corollary}
\emph{proof:}
Applying $\alpha=2$ and neglecting the noise term in the dominator of~\eqref{eq:outage U1:TZF:TOTAL1} and then using the integral identity~\cite[Eq. (2.33.11)]{Integral:Series:Ryzhik:1992} we obtain~\eqref{eq:outage U1:TZF:TOTAL4_alpha2}. $\blacksquare$
\section{Cooperative NOMA With NNNF Selection }
In this section, we investigate the outage performance of NNNF user selection
where the users' CSI are utilized to select the near and far users with the shortest distance to the BS. Accordingly, NNNF strategy can maximize the outage performance of both the near and far users.
\subsection{Outage Probability of the Near Users}
By invoking~\eqref{eq:outage event near}, we can characterize the outage probability of the near users. We have the following key result:
\begin{proposition}
\label{prob:outage:u1:TZF:nnnf}
The outage probability of $\SUun$ with the TZF scheme is given by
\vspace{-0.1em}
\begin{align}\label{prob:outage:u1:TZF:nnnf}
&\PoutuTZFn=\frac{\upsilon_n}{2\pi}\int_{0}^{R_1}\int_{-{\pi}}^{{\pi}}r e^{-\pi\lambda_nr^2}\left(1-\right.\\&\left.\frac{e^{-\mu r^\alpha}}{1\!+\!q_r\rho_r\mu\left(R_1^2\!+\!r^2-\!2rR_1cos(\theta_r\!-\!\theta_i)\right)^{-\frac{\alpha}{2}}r^{\alpha}}
\right) d\theta_i dr,\nonumber
\end{align}
where $\upsilon_n=\frac{2\pi\lambda_n}{1-e^{-\pi\lambda_nR_1^2}}$.
\end{proposition}
\emph{proof:}
Similar to~\eqref{eq:out:u1:tzf:def2}, $\PoutuTZFn$ for $\SUun$ can be written as
\vspace{-0.5em}
\begin{align}\label{eq:out:u1:def2:nnnf}
\PoutuTZFn=&\Prob\left(Y_1\leq {\left(\rho_r\ell(\SUr,\SUun)^{-\alpha}Y_0+1\right)\mu \ell(\SUun)^{\alpha}}\big|\right.\nonumber\\&\left. Y_0, N_{U_1}\geq1\right),
\end{align}
where $\ell(\SUun)$ is the distance from the nearest $\SUun$ to the BS and $\ell(\SUr,\SUun)$ is the distance from $\SUr$ to the nearest $\SUun$.
Following similar lines as in the derivation of~\eqref{eq:out:u1:def3}, $\PoutuTZFn$ for $\SUun$ can be written as
\vspace{-0.5em}
\begin{align}\label{eq:out:u1:nnnf}
&\PoutuTZFn =\frac{1}{2\pi}\int_{0}^{R_1}\int_{-\pi}^{\pi}f_{n^*}(r)\left(1\!-\!\right.\\&\left.\frac{e^{-\mu r^\alpha}}{1\!+q_r\!\rho_r\mu\left(R_1^2\!+\!r^2\!-\!2rR_1cos(\theta_r\!-\!\theta_i)\right)^{-\frac{\alpha}{2}}r^{\alpha}}
\right) d\theta_i dr,\nonumber
\end{align}
where $f_{n^*}(r)$ is the pdf of the shortest distance from $\SUun$ to the BS which is given by~\cite{Zhiguo:JSAC:2016}
$f_{n^*}(r)=\upsilon_n r e^{-\pi\lambda_nr^2}$. Substituting $f_{n^*}(r)$ into~\eqref{eq:out:u1:nnnf} we obtain the outage probability in~\eqref{prob:outage:u1:TZF:nnnf}. $\blacksquare$
\begin{Corollary}
The outage probability of $\SUun$ of NNNF with the TZF scheme can be approximately upper bounded ($\eta=1$) (lower bounded ($\eta=-1$)) as
\vspace{-0.5em}
\begin{align}\label{eq:out_tzf_near asymp_upper_NNNF}
&\PoutuTZFUn\approx\frac{\pi\upsilon_nR_1}{2M}\sum_{m=1}^{M}\sqrt{(1-\phi_m^2)}\left(1-\right.\\&~\left.\frac{e^{-\mu c_m^\alpha}}{1+q_r\rho_r\mu c_m^{\alpha}\left(R_1^2+c_m^2-2\eta R_1c_m\right)^{-\frac{\alpha}{2}}}\right)c_m e^{-\pi\lambda_n c_m^2}.\nonumber
\end{align}
\end{Corollary}
\emph{proof:}
The proof is similar to Corollary~\ref{cor:upp:outage:u1:TZF}. $\blacksquare$
\subsection{Outage Probability of the Far Users}
Using the definition in~\eqref{eq:outage u2}, we  analyze the outage
probability of the far users. The following proposition presents the outage probability valid for an arbitrary  $\alpha$.
\begin{proposition}
\label{prob:outage:u2:TZF}
The outage probability of $\SUuun$ with the TZF scheme is given by~\eqref{eq:outage U1:TZF:TOTAL:nnnf} at the top of the page
where $b_4=\frac{\upsilon_f \pi (R_3-R_2) e^{\pi\lambda_fR_2^2}}{2M}$ and $s_m=\frac{R_3-R_2}{2}(\phi_m+1)+R_2$.
\end{proposition}
\emph{proof:}
The outage probability of $\SUuun$ can be expressed as
\setcounter{equation}{31}
\begin{align}\label{eq:outage U1:TZF:NNNF:TOTAL2}
\PoutuuTZFn=&\Prob\left(\frac{a_{2,i}Y_2}{a_{1,i}Y_2+1}<\tau_2|N_{U_2}\geq1\right)+\nonumber\\
&\Prob\left(\frac{a_{2,i}Y_2}{a_{1,i}Y_2+1}>\tau_2|N_{U_2}\geq1\right)\times\nonumber\\
&\Prob\left(\rho_r\ell(\SUr,\SUuun)^{-\alpha}Y_3 <\tau_2|N_{U_2}\geq1\right).
\end{align}
Since $R_2\gg R_1$, we can approximate $\ell(\SUr,\SUuun)\approx\ell(\SUuun)$. Similar to~\eqref{eq:outage U1:TZF:TOTAL2},  $\PoutuuTZFn$ can be evaluated as
\vspace{-0.5em}
\begin{align}\label{eq:outage U1:TZF:TOTAL2_nnnf}
&\PoutuuTZFn
=1-\frac{1}{\Gamma(\Nrx)}\Gamma\!\left(\Nrx,\frac{\tau_2R_1^\alpha}{\rho_s(a_{2,i}-\tau_2a_{1,i})}\right)\!+\!\frac{1}{\Gamma(\Nrx)}\!\times\!\nonumber\\
&\Gamma\!\left(\Nrx\!,\frac{\tau_2R_1^\alpha}{\rho_s(a_{2,i}\!-\!\tau_2a_{1,i})}\right)
\Prob \left(Y_3\!<\!\frac{\tau_2}{\rho_r \ell(\SUuun)^{-\alpha}}|  N_{U_2}\geq1\right)\!.
\end{align}
The RV, $Y_3$ follows the chi-square distribution with $2(\Ntx-1)$ DoF and thus
\vspace{-0.5em}
\begin{align}\label{eq:outage U1:TZF:TOTAL3_nnnf}
F_{Y_3}\left(\beta \ell(\SUuun)^{\alpha}\right)&=\int_{R_2}^{R_3}\left(1\!-\!e^{-\beta r^\alpha}\sum_{k=0}^{\Ntx\!-\!2}\frac{\beta^{k}}{k!} r^{\alpha k}\right)f_f^*(r) dr,
\end{align}
where $f_f^*(r)$ is the pdf of the nearest $\SUuun$ which is given by~\cite{Zhiguo:JSAC:2016}
\vspace{-0.5em}
\begin{align}\label{eq:f_f}
f_f^*(r)=\upsilon_f r e^{-\pi\lambda_{f}(r^2-R_2^2)},
\end{align}
where $\upsilon_f=\frac{2\pi\lambda_f}{1-e^{-\pi\lambda_f(R_3^2-R_2^2)}}$. Substituting~\eqref{eq:f_f} into~\eqref{eq:outage U1:TZF:TOTAL3_nnnf}, we obtain
\vspace{-0.5em}
\begin{align}\label{eq:outage U1:TZF:TOTAL4_nnnf}
F_{Y_3}\left(\beta \ell(\SUuun)^{\alpha}\right)&=1-\upsilon_f e^{\pi\lambda_fR_2^2}\sum_{k=0}^{\Ntx-2}\frac{\beta^{k}}{k!}\Psi,
\end{align}
where $\Psi=\int_{R_2}^{R_3}r^{\alpha k+1}e^{-(\beta r^{\alpha}+\pi\lambda_f r^2)}dr$. An exact evaluation of $\Psi$ is mathematically intractable. Motivated by this, we use Gaussian-Chebyshev quadrature to find an approximation as
\vspace{-0.5em}
\begin{align}\label{eq:Psi}
\Psi\!\approx\frac{\pi(R_3-R_2)}{2M}\left(\sum_{m=1}^{M}\!\sqrt{1-\phi_m^2}s_m^{\alpha k+1}e^{-\!\left(\beta s_m^\alpha\!+\!\pi \lambda_f s_m^2 \!\right)}\right).
\end{align}
To this end,  by substituting~\eqref{eq:Psi} into~\eqref{eq:outage U1:TZF:TOTAL4_nnnf} and then substituting the result into~\eqref{eq:outage U1:TZF:TOTAL2_nnnf} we get the desired result.
$\blacksquare$
\begin{Corollary}
\label{prob:outage:u2:TZF_alpha2_nnnf}
In the high SNR regime and for the special case $\alpha=2$, the outage probability of $\SUuun$ of NNNF  can be simplified  as
\vspace{-0.5em}
\begin{align}\label{eq:outageU2:TZF:alpha2_nnnf}
\PoutuuTZFn=
 {1-\frac{b_5}{2}\sum_{k=0}^{\Ntx-2}\beta^{k}\left(H(R_2)-H(R_3)\right)},
\end{align}
where $H(x)=e^{-\delta x^2}\sum_{j=0}^{k}\frac{x^{2j}}{j!\delta^{k+1-j}}$, $\delta=\beta+\pi \lambda_f$ and $b_5=\upsilon_f e^{\pi\lambda_fR_2^2}$.
\end{Corollary}
\emph{proof:}
Applying $\alpha=2$ in~\eqref{eq:outage U1:TZF:TOTAL4_nnnf} and using the integral identity~\cite[Eq. (2.33.11)]{Integral:Series:Ryzhik:1992} we can obtain~\eqref{eq:outageU2:TZF:alpha2_nnnf}. $\blacksquare$
\begin{figure}
\centering
\includegraphics[width=79mm, height=60mm]{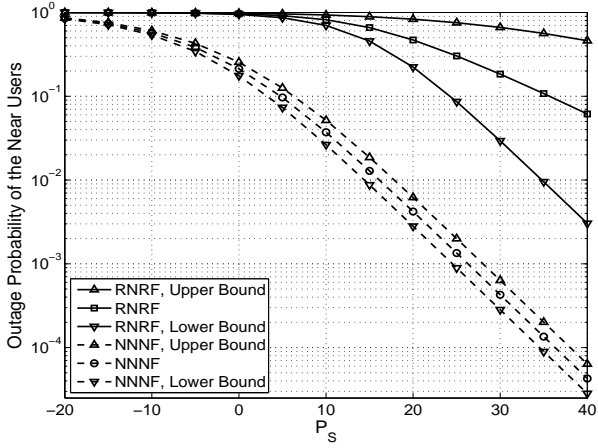}
\vspace{-0.4em}\caption{Outage probability of the near users  versus $P_S$ (in dBW). The derived upper bounds and lower bounds for both strategies are also presented where $\alpha=3$, $P_R=40$ dBW,  $\mathcal{R}_1=\mathcal{R}_2=0.4$ bps/Hz, and $\lambda_n=10$. }
\vspace{-.8em}
\label{fig: outage_Psource_Near}
\end{figure}

\begin{figure}
\vspace{-.7em}
\centering
\includegraphics[width=82mm, height=62mm]{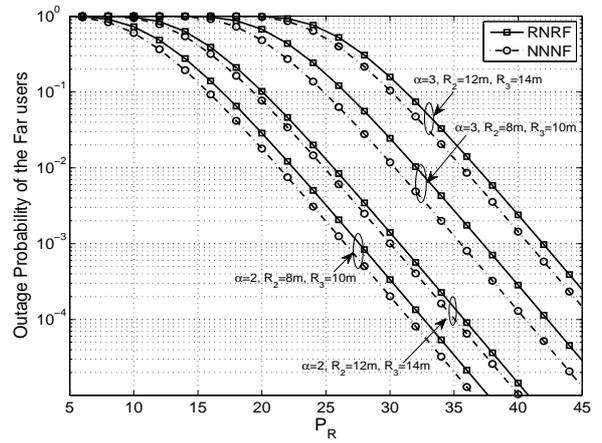}
\vspace{-0.4em}\caption{Outage probability of the far users versus $P_R$ (in dBW) for different $\alpha$, $R_2$ and $R_3$ where $R_1=2$ m, $N_T=N_R=3$, $\mathcal{R}_1=\mathcal{R}_2=0.4$ bps/Hz, $\lambda_n=\lambda_f=10$, and $P_S=30$ dBW.}
\vspace{-.8em}
\label{fig: outage_Psource_Far}
\end{figure}

\section{Numerical Results and Discussion}
n this section, we  present numerical results to validate our analysis, demonstrate the performance and investigate the impact of key system parameters. We set $a_1=0.2$, $a_2=0.8$, $N_0=1$ dBW and $K=3$.

Fig.~\ref{fig: outage_Psource_Near} shows the outage probability of the near users versus $P_S$ for RNRF and NNNF strategies, respectively.
Numerical results evaluated from~\eqref{prob:outage:u1:TZF} and~\eqref{prob:outage:u1:TZF:nnnf} were in agreement with simulations not shown
and NNNF upper and lower bounds are tight. This is because, in the NNNF strategy, the distance of the nearest user to the BS, i.e., $\ell(\SUun)$, approaches to zero  and hence the term $2R_1\ell(\SUun)\cos(\theta_r-\theta_{i})$ in $\ell(\SUr,\SUun)=\sqrt{R_1^2+\ell(\SUun)^2-2R_1\ell(\SUun)\cos(\theta_r-\theta_{i})}$ is small which makes the difference of the bounds and the exact values insignificant.

Fig.~\ref{fig: outage_Psource_Far} compares the outage probability of the far users versus $P_R$ with different $\alpha$, $R_2$, and $R_3$.  It is observed that NNNF exhibits a superior outage performance in comparison to the RNRF strategy.  As expected, the outage probability increases as $\alpha$ increases. In addition, we  see that  increasing the radius of the far user's ring deteriorates the outage performance of the NNNF and RNRF strategies.

\begin{figure}
\centering
\includegraphics[width=82mm, height=64mm]{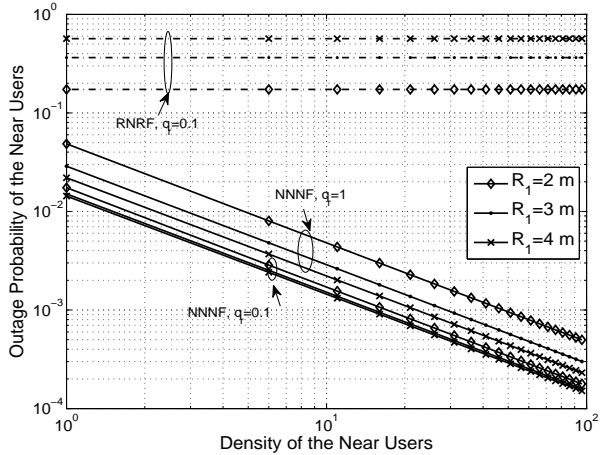}
\vspace{-0 em}\caption{Outage probability of the near users versus density of the near users for different strength of inter-user interference $q_r$ and  $R_1$ where $\alpha=3$,  $P_S=P_R=10$ dBW,  and $\mathcal{R}_1=0.4$ bps/Hz.}
\vspace{-.6em}
\label{fig: outage_density_Near}
\end{figure}

Fig.~\ref{fig: outage_density_Near} shows the outage performance of the near users versus the density $\lambda_n$ for different levels of the inter-user interference strength at the near users, $q_r$, and different $R_1$. It can be observed that the outage performance of  RNRF is independent of the $\lambda_n$. This is because of the fact that RNRF selects users randomly and hence increasing number of near users will not affect its performance. On the other hand, the outage probability of NNNF decreases when $\lambda_n$ increases. It is intuitive since a higher number of near users offers better near user positions, and thus better outage performance could be expected. As expected, the outage probability decreases as $q_r$ decreases. Also, the outage probability of RNRF increases by increasing $R_1$. However, NNNF presents lower outage probability for higher $R_1$. This is more noticeable at high values of $q_r$ and the difference between the curves are negligible for low $q_r$. In the NNNF strategy the nearest user to the BS is selected as a near user where increasing $R_1$ will not change its position notably. On the other hand, the outage probability of the near users degrades due to interference from the relay to the near users which increases as the $R_1$ decreases. However, when $q_r$ is sufficiently small, the effect of this interference becomes invisible.

\begin{figure}
\centering
\includegraphics[width=80mm, height=60mm]{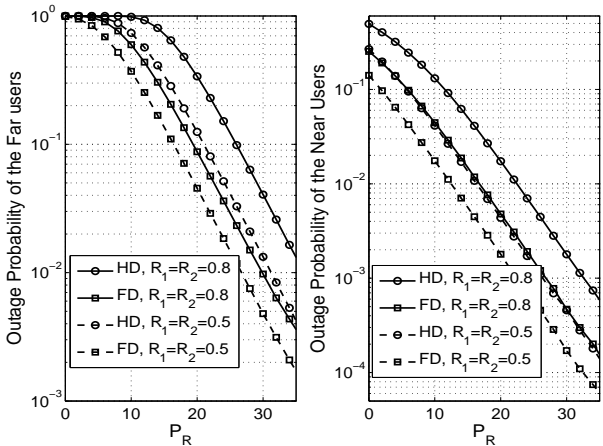}
\vspace{-0 em}\caption{Outage probability comparison between FD and HD for  NNNF strategy versus $P_R$ (in dBW) for different user rates (bps/Hz) where $P_s=40$ dBW for the far users and $20$ dBW for the near users, $\alpha=3$, $R_1=2$ m, $R_2=8$ m, $R_3=10$ m, $N_T=3$, $N_R=1$, and $\lambda_n=\lambda_f=1$.}
\vspace{-.8em}
\label{fig: outage_Praly_FDHD}
\end{figure}

Fig.~\ref{fig: outage_Praly_FDHD} shows the outage probability of the near and far users for FD and HD schemes with different target rates under the ``RF chain preserved'' condition~\cite{Himal:2014:JSAC}. The outage performance of FD cooperative NOMA with TZF outperforms HD cooperative NOMA system. The main reason is that FD cooperative NOMA with TZF can recover the spectral loss incurred by conventional HD relaying and cancel the residual SI at the relay at the same time. Fig.~\ref{fig: outage_Praly_FDHD} also illustrates that increasing the target rates deteriorates the
outage performance of both schemes since it raises the decoding threshold.

\section{conclusion}
In this paper, we have investigated a NOMA system assisted by an FD multi-antenna relay where MRC and ZF processing are used as receive and transmit beamformers, respectively. Considering a realistic scenario with imperfect interference cancellation and using tools from stochastic geometry, exact analytical expressions as well as closed-form upper bound and lower bound for the outage probability of the near users were derived for RNRF and NNNF user selection strategies. We also presented numerical results to validate our analysis. Our results show that even with imperfect inter-user interference cancellation at the near users, FD transmissions with TZF could achieve a lower outage probability as compered to HD transmissions. In addition, we found that improving the outage performance can be achieved by reducing the user's zone or increasing the density of the users.
\balance
\bibliographystyle{IEEEtran}

\end{document}